# The long-term scientific benefits of a space economy

Ian A. Crawford

*Department of Earth and Planetary Sciences, Birkbeck College, University of London, London, United Kingdom.*

*Email address: i.crawford@bbk.ac.uk*

**ABSTRACT**

Utilisation of the material and energy resources of the Solar System will be essential for the development of a sustainable space economy and associated infrastructure. Science will be a major beneficiary of a space economy, even if its major elements (e.g. space tourism, resource extraction activities on the Moon or asteroids, and large-scale in-space construction capabilities) are not developed with science primarily in mind. Examples of scientific activities that would be facilitated by the development of space infrastructure include the construction of large space telescopes, ambitious space missions (including human missions) to the outer Solar System, and the establishment of scientific research stations on the Moon and Mars (and perhaps elsewhere). In the more distant future, an important scientific application of a well-developed space infrastructure may be the construction of interstellar space probes for the exploration of planets around nearby stars.

**1. Introduction**

There is no doubt that science has been a major beneficiary of the space age. Ever since the discovery of Earth's radiation belts by only the second and third artificial Earth satellites ever launched (i.e. Sputnik 2 in 1957 and Explorer-1 the following year), scientific knowledge has poured down to Earth from spacecraft operating throughout the Solar System. Major scientific disciplines like astrophysics and planetary science have undergone paradigm-changing revolutions as a consequence, in the former case by the ability to conduct astronomical observations above the obscuring effects of Earth's atmosphere, and in the latter by being able to make *in situ* measurements of planets and other solar system bodies that previously could only be observed remotely using telescopes. In the second decade of the twenty-first century, it is sobering to contemplate just how limited our knowledge of the Universe would still be had the space age not begun when it did.

It is clear that science still has much to gain from continued access to space, but already there are concerns that access on the scale required to maintain the current rate of scientific discovery, let alone to increase it, may not be achievable with current funding models [1,2]. The main concern is that, as space missions become more complex, their cost grows faster than do scientific budgets, or even the Gross National Products of participating countries. As pointed out by Martin Elvis elsewhere in this volume [2], ultimately this must result in space science hitting a 'funding wall' which will curtail future growth, and thus limit future scientific discoveries. One way, and perhaps the only way, to avoid this funding limit will be to 'piggyback' space science and exploration on the activities of a future space-based economy developed largely for commercial purposes [e.g. 1-4].

## 2. Space resources

To be sustainable, any future space-based economy, whether built around commercial satellite operations, space tourism, mining the Moon and asteroids, or any combination of these, will increasingly rely on utilising the energy and material resources of the solar system [e.g. 4-9]. This is simply a consequence of the energy cost of lifting the required materials out of Earth's gravity well: if materials are to be used in space it will always be more economically attractive to source them in space. However, the process will be gradual and iterative, because many of these space resources will only be useful once an infrastructure exists to access and extract them.

There are at least four, mutually reinforcing, ways in which the scientific exploration of space will benefit from the development of a space economy built around the utilisation of extraterrestrial resources.

*2.1 Economic benefits of using space resources in space to build, provision, and maintain scientific instruments and outposts*

It has long been recognized that future space exploration activities would benefit from utilising extraterrestrial resources wherever possible (an application known as *In Situ* Resource Utilisation, or ISRU), as this would avoid having to lift them out of Earth's gravity [4, 7, 10-12]. For example, scientific outposts on the Moon and Mars would benefit from using indigenous water resources (e.g. for drinking, personal hygiene, and as a source of both hydrogen and oxygen). Similarly, future space stations, satellites (including, in the present context, the next generation of large space telescopes), and space probes to the outer solar system would benefit if the hydrogen and oxygen needed for rocket fuel (as well as oxygen to breath if human crews are involved) could be sourced in space (e.g. from the lunar poles [8,13] or from hydrated asteroids [14,15]). For such applications, it is not the intrinsic value of the resources themselves that are scientifically enabling but the economic savings resulting

from reduced launch masses from Earth. As Drmola and Mareš [16] succinctly put the case for ISRU:

> "It is not the prospect of procuring something we covet here on the surface of the Earth that makes this venture attractive, but rather the idea of not having to wage an expensive battle with Earth's gravity each time we want to make use of something as mundane as water in space."

2.2 *Scientific discoveries made in the course of resource prospecting/extraction*

Companies engaged in prospecting for exploitable raw materials elsewhere in the solar system will inevitably rely on the knowledge and technical expertise of the planetary science community. However, by the same token, planetary science will learn much from this prospecting activity owing to improved access to a wider range solar system bodies (e.g. comets, asteroids, and lunar and planetary surfaces) than would otherwise occur (see the discussion summarised in Ref. [4]). This will greatly increase opportunities for making *in situ* measurements, and for returning a diverse range of samples to Earth for analysis, all of which will increase our understanding of the origin and evolution of the Solar System. There is a clear analogy here with the symbiotic relationship which exists between the geological sciences and the resource extraction industries on Earth – the latter needs the expertise of the former in order to locate economically exploitable resources, but the former also benefits from discoveries made, and techniques developed, by the latter. The same dynamic is sure to play out in space as the economic development of the solar system proceeds.

2.3 *Leveraging the growth of the space economy to pay for space science*

Ultimately the 'New Space' entrepreneurs behind the commercialisation of space activities are seeking to make a profit, and the bigger the space economy becomes the larger will be the profits generated. Some of these profits will doubtless be re-invested in further space exploration and development, leading to increased opportunities for scientific discovery as outlined in Section 2.2 above. However, it will presumably also be the case that companies profiting from commercial space activities will be subject to taxation by some Earth-bound jurisdiction (whether this be national governments or some as yet to be implemented international authority [e.g. 5,17-19]). Thus, by instituting creative regulatory mechanisms, it may be possible to arrange for some of the profits generated in space to be invested in government-led 'blue skies' space science activities unrelated to the commercial activities themselves. Financing, or part financing, of the next generation of large space telescopes, or human missions to Mars, might be examples. In any case, it seems clear that once space activities begin to pay for themselves, and clearly contribute to the global economy rather than being (or being perceived to be) a drain upon it, then more opportunities to finance purely scientific endeavours in space are likely to arise.

2.4 *Utilisation of the infrastructure developed to support commercial space activities*

Last, but not least, space science and exploration will benefit from the *infrastructure* developed to support a space economy. Consistent with its usual definition, by space infrastructure I mean all the transportation capabilities, habitats, trained personnel, and other capabilities that will facilitate wide-ranging human operations in the space environment (Fig. 1). It is inherent in the nature of infrastructure that it can support multiple activities; for example, a spacecraft designed to transport people could carry tourists, asteroid miners, and, in the present context, scientists all at the same time.

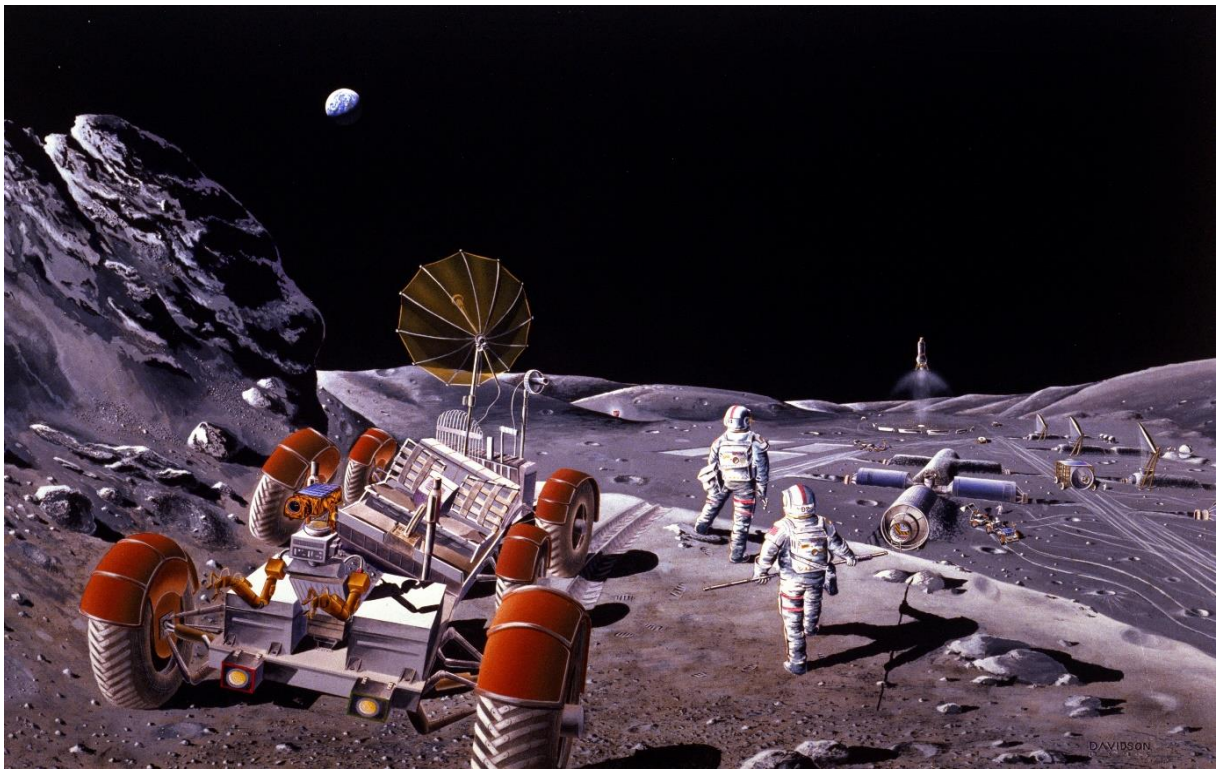

**Fig. 1**. Infrastructure on the Moon. This artist's drawing of future lunar exploration includes multiple infrastructural elements, including habitats, local transportation (rover), interplanetary transportation (rocket), and trained personnel (astronauts). Such an infrastructure could support multiple activities, including resource prospecting, tourism, and, in the present context, scientific research. Sustainable operation of such an infrastructure will require the use of *in situ* resources and will enable planetary exploration on a scale that will not otherwise be possible (NASA).

As I have noted elsewhere [20], there is a clear analogy here with the multiple uses of civil aviation infrastructure on Earth. For example, astronomers, geologists and zoologists invariably make use of an extensive commercial aviation infrastructure (which, of course, is largely underpinned by the tourism industry) in order to visit their observatories and field localities, without having to design, build, and operate commercial airliners. This is not a trivial point: because scientific budgets would be wholly inadequate to develop such an infrastructure, if it had not been created for other reasons then a lot of scientific activity on Earth simply could not occur. Future scientists operating on the Moon, Mars and asteroids (and indeed throughout the solar system) would similarly benefit from a commercial interplanetary transportation infrastructure. Similar comparisons can be made with other aspects of infrastructure. For example, just as telescope construction on Earth relies on the availability of a commercial construction industry, so future large space telescopes will be enabled by the existence of a space-based equivalent, even if most of its business case relies on commercial activities, such as building orbital hotels or satellite solar power stations, quite unrelated to science.

## 3. Thinking further ahead

In the above, I have alluded to relatively near-term (say within the next fifty years) examples of scientific benefits likely to result from the gradual development of a space economy and associated infrastructure. These include the construction of large space telescopes, in depth studies of a wide range of asteroids, cheaper and more capable robotic missions to the outer planets, and the establishment of scientific research stations on the Moon and, in due course, Mars. However, once a space economy is established it may be expected to grow exponentially, and in time this will enable still more ambitious feats of scientific exploration.

One such longer term opportunity might be human exploration missions to the giant planets and their moons. Currently, it is taken for granted that missions to the outer solar system will, of necessity, be robotic, mainly owing to the very long travel times required (with attendant consequences for life support consumables, radiation shielding, and microgravity mitigation), and the very hazardous radiation environments in the vicinity of the giant planets. However, if it were ever possible to send human explorers to, say, Europa or Titan, then all the advantages that human explorers will have over robots when exploring the Moon or Mars (i.e. greater versatility, on-the-spot decision making, intelligent sample collection, maintenance and deployment of complex equipment, greater opportunities for serendipitous discovery, etc, [21-23]) would apply to those bodies as well. Given the scientific importance of the icy worlds of the outer solar system, not least from an astrobiological perspective [24,25], the scientific rewards from being able to send people there (or at least to their vicinity so as to permit real-time telerobotic operations [26]) could be significant.

In spite of the well-recognized difficulties, studies of human missions to the outer solar system have been conducted [e.g. 27] and, although technically demanding, are certainly not impossible. Indeed, most of the seemingly insurmountable difficulties arise from the large spacecraft mass that would needed for any human mission to the outer solar system. Such missions are likely to require launch masses of several thousand tonnes, driven largely by the mass of the propulsion system (presumably nuclear of some kind) needed to reduce transit times to acceptable levels (i.e. months to a small number of years), the mass of life support consumables required, and the mass of radiation shielding [27]. However, while probably unrealistic if launched directly from Earth, the economics of such a mission would be completely transformed in the context of a space economy where the spacecraft could be assembled and launched, and many of the consumables sourced, from low gravity environments in the inner solar system. It will of course take time (surely many decades) for a solar system economy to 'bootstrap' itself up to a level at which it could realistically support such activities, but when it is able to do so the scientific exploration of the outer solar system is likely to be a major beneficiary.

In the more distant future, an important scientific benefit of a well-developed space infrastructure may be the construction of interstellar space probes for the exploration of planets around nearby stars. The history of planetary exploration clearly shows that *in situ* investigations by space probes are required if we are to learn about the interior structures, geological evolution, and possible habitability of the planets in our own solar system, and so it seems clear that spacecraft will eventually be needed for the investigation of other planetary systems as well [28]. For example, if future astronomical observations from the solar system (perhaps using large space telescopes themselves built and paid for using space resources) find evidence suggesting that life might exist on a planet orbiting a nearby star, *in situ* measurements will probably be required to get definitive proof of its existence and to learn more about its underlying biochemistry, ecology, and evolutionary history. This in turn will eventually require transporting sophisticated scientific instruments across interstellar space.

However, the scale of such an undertaking should not be underestimated. Although very low-mass laser-pushed nano-craft, such as are being considered by Project Starshot [29], could conceivably be launched directly from Earth, the scientific capabilities of such small payloads will surely be very limited. In order to get a scientifically useful payload (which, even allowing for future progress in miniaturisation, will surely need to have a mass of at least several tonnes, and possibly very much more if detailed investigations are to be made [30]) to even the nearest stars (all several light-years away) within scientifically useful timescales (say ≤100 years) then spacecraft velocities of order 10% of the speed of light will be required. Of course, this far exceeds present capabilities and will likely require vehicles of such a size, with such highly energetic (and thus potentially dangerous) propulsion systems [e.g. 31-33], that their construction and launch will surely have to take place in space. The potential long-term scientific benefits of an interstellar spacefaring capability are hard to exaggerate, but it seems certain that it is a capability that will only become possible in the context of a well-developed space economy with access to the material and energy resources of our own solar system [5,31].

## 4. Conclusions

The solar system is rich in energy and material resources which could potentially support a vibrant future space economy. Although many aspects of this economic activity will probably be pursued for purely commercial reasons (e.g. space tourism, and the mining of the Moon and asteroids for economically valuable materials), science will nevertheless be a major beneficiary. Indeed, science will benefit from all stages in the 'bootstrapping' of a space economy, from initial prospecting activities on the Moon and asteroids, through to the utilisation of the resulting resources to expand space activities. In particular, science will benefit from the *infrastructure* developed to support a space economy, which will help facilitate the construction of large space telescopes, the establishment of scientific research stations on the Moon and Mars (and perhaps elsewhere), and the mounting of ambitious space missions (including human missions) to the outer Solar System. In the more distant future, an important scientific benefit of a well-developed space infrastructure may be the construction of interstellar space probes for the exploration of planets around nearby stars.